\documentclass[sigconf]{acmart}
\AtBeginDocument{%
  }

\usepackage{tcolorbox}
\usepackage{xcolor}
\usepackage{graphicx}
\usepackage{url}
\usepackage{colortbl}
\usepackage{array}
\usepackage{tikz}
\usetikzlibrary{positioning,arrows.meta,shapes.geometric,fit,backgrounds}
\definecolor{rulered}{RGB}{229,71,86}
\definecolor{rulepink}{RGB}{253,234,235}
\definecolor{rulegray}{RGB}{231,231,231}

\newtheorem{definition}{Definition}

\usepackage{enumitem}
\definecolor{todored}{RGB}{180,30,30}

\newcommand{\leaff}{\mathrm{leaf}}

\newcommand{\Dset}{\mathcal{D}}
\setcopyright{acmlicensed}
\copyrightyear{2026}
\acmYear{2026}
\acmDOI{XXXXXXX.XXXXXXX}
\acmConference[KDD CJ '26]{5th End-to-End Customer Journey Workshop at KDD 2026}{August 10, 2026}{Jeju Island, South Korea}
\acmISBN{978-1-4503-XXXX-X/XXXX/XX}

\begin{document}

\title{Algorithmic Contract Design at Scale: Adaptive Peer Comparison for Enterprise Pricing}

\author{Jason Huang}
\affiliation{%
  \institution{Databricks}
  \city{New York}
  \state{NY}
  \country{USA}
}
\email{jason.huang@databricks.com}

\author{Song (Vinson) Wei}
\authornote{Both authors contributed equally to this research.}
\affiliation{%
  \institution{Databricks}
  \city{Mountain View}
  \state{CA}
  \country{USA}
}
\email{vinson.wei@databricks.com}

\renewcommand{\shortauthors}{Huang et al.}

\begin{abstract}
In enterprise software, a contract commits the customer to a usage volume over a fixed term in exchange for discounted pricing. These contracts are individually negotiated across many dimensions---size, duration, industry, product mix, usage history---and without a data-driven reference point, discounts tend to be overly generous. Manual governance review enforces discipline but at days-scale per contract, with inconsistency across reviewers and no real-time feedback to sellers. We present \emph{Contract Scoring}, a peer-based grading system deployed on every contract at Databricks. The system identifies empirically similar historical contracts via adaptive nearest neighbors over ensemble trees, where shared leaf membership defines a data-driven similarity learned from the discount target. It returns a letter grade with per-product-line breakdown in seconds; the underlying peer set is available to the centralized review team for audit. Sellers treat the grade as a contract design ``exit criterion'', iteratively adjusting discount structures until the grade reflects their intended tradeoff. Deployment evidence shows measurable discount discipline across the scored portfolio, with a commercially significant impact on revenue.
\end{abstract}



\maketitle

\section{Introduction}

Discount governance in enterprise software is a universal challenge: contracts are individually negotiated, and without a data-driven reference point, discounts tend to be overly generous~\cite{Phillips2005}. Two standard approaches---static rule tables and manual desk review---each fail differently. Static rules, parameterized only by contract size and duration, cannot capture the dimensions that drive discount appropriateness (e.g., industry, product mix, usage history, customer tenure, etc.). Manual review addresses the remainder but at days-scale per contract, with inconsistency across reviewers and no real-time feedback to enable efficient contract revision.

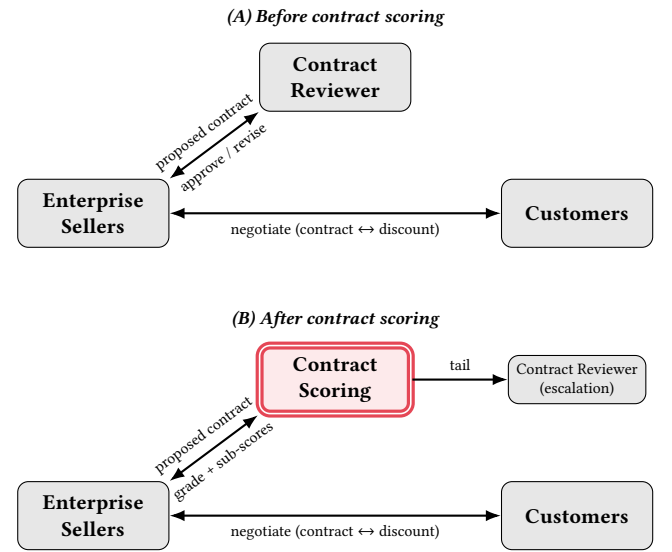
\begin{figure}[!h]
\centering
\begin{tikzpicture}[
  party/.style={draw, rounded corners, minimum height=9mm, minimum width=20mm,
                align=center, font=\small\bfseries, fill=rulegray},
  reviewer/.style={party, fill=rulepink, draw=rulered, thick},
  scoring/.style={party, fill=rulepink, draw=rulered, very thick, double},
  smallnode/.style={draw, rounded corners, minimum height=6mm, minimum width=18mm,
                    align=center, font=\scriptsize, fill=rulegray},
  flow/.style={-{Latex[length=2mm]}, thick},
  bidir/.style={{Latex[length=2mm]}-{Latex[length=2mm]}, thick},
  panel/.style={font=\footnotesize\itshape\bfseries}
]
\node[party] (sellerA) at (0,0) {Enterprise\\Sellers};
\node[party] (custA) at (6.4,0) {Customers};
\node[party] (revA) at (3.2,1.8) {Contract\\Reviewer};
\draw[bidir] (sellerA.east) -- node[below,font=\scriptsize]{negotiate (contract $\leftrightarrow$ discount)} (custA.west);
\draw[bidir] (sellerA.north east) -- node[sloped,above,font=\scriptsize]{proposed contract} node[sloped,below,font=\scriptsize]{approve / revise} (revA.south west);
\node[panel, above=1mm of revA] {(A) Before contract scoring};

\node[party] (sellerB) at (0,-4.0) {Enterprise\\Sellers};
\node[party] (custB) at (6.4,-4.0) {Customers};
\node[scoring] (scoreB) at (3.2,-2.2) {Contract\\Scoring};
\node[smallnode] (revB) at (6.4,-2.2) {Contract Reviewer\\(escalation)};
\draw[bidir] (sellerB.east) -- node[below,font=\scriptsize]{negotiate (contract $\leftrightarrow$ discount)} (custB.west);
\draw[bidir] (sellerB.north east) -- node[sloped,above,font=\scriptsize]{proposed contract} node[sloped,below,font=\scriptsize]{grade + sub-scores} (scoreB.south west);
\draw[flow] (scoreB.east) -- node[above,font=\scriptsize]{tail} (revB.west);
\node[panel, above=1mm of scoreB] {(B) After contract scoring};
\end{tikzpicture}
\vspace{-0.05in}
\caption{Three-party dynamic. \textbf{(A)}~Before: a Contract Reviewer gates every contract at days-scale. \textbf{(B)}~After: Contract Scoring returns a grade at machine speed; human review is reserved for the complex tail.}
\label{fig:three-party-before-after}
\end{figure}

\noindent\paragraph{Three parties, misaligned incentives.} Enterprise contract pricing involves three parties with conflicting objectives (Figure~\ref{fig:three-party-before-after}): \textbf{Customers} seek the lowest effective price, and \textbf{enterprise sellers} are incentivized to close contracts, hence the sellers naturally accommodate customer pressure by offering generous discounts; \textbf{Contract reviewers} enforce discount discipline on behalf of the firm---but this manual gate takes days per contract, provides no real-time feedback to sellers, and is inconsistent across reviewers. At Databricks, we present \emph{Contract Scoring} to address this challenge: a peer-based grading system integrated into the sales workflow that returns a letter grade and per-product-line breakdown in real time. Human reviewers remain in the loop for the genuinely complex tail (contracts with non-standard terms or requiring executive approval); for all proposed contracts, the seller receives immediate feedback on how the proposed discount compares to similar historical contracts.

\vspace{-0.1in}
\vspace{-0.1in}

\noindent\paragraph{Why peer-based grading?} Prescriptive pricing optimization~\cite{Colias2023,Biggs2021} addresses a related problem but requires demand-curve estimation---modeling how acceptance probability changes with discount level. Under the contract negotiation context, acceptance depends on latent factors---relationship history, competitive pressure, strategic priorities---that are difficult to observe and rarely captured in available data. Even where feasible, a recommended price is opaque: \citet{Biggs2021} note that interpretability is a key adoption barrier for ML-driven pricing. Classical $k$-nearest-neighbor ($k$NN) retrieval in raw feature space requires a hand-picked distance metric and degrades under the curse of dimensionality; quantile regression forests~\cite{Meinshausen2006} lose the explicit peer narrative that makes the grade defensible to stakeholders. Peer-based grading avoids these limitations. Peers map naturally onto how the business reasons (i.e., ``contracts like this one''); the peer ranking is auditable by the review team; the peer set provides the full conditional distribution rather than a point estimate; and real-time response turns the grading system into a navigable design API. Our peer-construction primitive builds on the adaptive $k$NN interpretation of ensemble trees~\cite{LinJeon2006}, extended to distributional settings~\cite{Meinshausen2006,Cevid2022}. Comparable-firm valuation in finance~\cite{BhojrajLee2002} and data-driven peer sets via text similarity~\cite{HobergPhillips2016} share the structural idea but define similarity by proxy rather than learning it from the discount target. See Appendix~\ref{sec:app-related-work} for an extended literature survey.

\noindent\paragraph{Contributions.} (1)~We frame contract-discount grading as adaptive $k$NN via ensemble trees (\S\ref{sec:method}). (2)~A production system providing real-time, transparent design feedback on every contract at Databricks (Figure~\ref{fig:architecture}). (3)~Empirical evidence that sellers respond to the grade by adjusting discounts toward the boundary: bunching~\cite{Kleven2016,Saez2010} at letter-grade boundaries, producing measurable discount discipline across the scored portfolio with a commercially meaningful impact on revenue (\S\ref{sec:impact}).

\section{Methodology}

We first formalize the grading problem and define the composite target variable, then show how ensemble trees solve the peer-construction step that makes grading feasible.

\subsection{Problem Setup}

Contract-discount grading requires identifying empirically similar contracts in a high-dimensional space where the target is a composite quantity and naive distance metrics are unreliable. Each contract is represented by a feature vector $x_i \in \mathcal{X}$, where $\mathcal{X}$ is a mixed continuous-categorical space spanning contract structure (e.g., size, duration), product-line usage patterns, and customer characteristics (e.g., tenure, new-business indicator), among others. The full production feature space is intentionally not enumerated here. A first contract and a renewal of comparable size occupy different regions of this space---each negotiation carries unique context, which is what makes case-by-case peer comparison essential here.

Each contract spans multiple product lines, each with its own discount. We grade the contract as a single object by computing a usage-weighted average across per-product-line discounts.
Let $\Dset = \{(x_i, y_i)\}_{i=1}^N$ denote the historical training set of $N$ contracts, where $y_i$ is the usage-weighted discount:
\begin{definition}[Usage-weighted discount]\label{def:uwd}
Let $J$ be the total number of product lines. For contract $i$, let $\delta_{i,j}$ be the discount for product line $j$ and $\omega_{i,j}$ a product-line weight, with $\omega_{i,j} = 0$ for product lines the contract does not include. The usage-weighted discount is:
\begin{equation}
y_i = \sum_{j=1}^{J} \omega_{i,j} \cdot \delta_{i,j}, \qquad \sum_{j=1}^{J} \omega_{i,j} = 1, \quad \omega_{i,j} \geq 0.
\end{equation}
\end{definition}

Given a query contract $(x_{\texttt{q}}, y_{\texttt{q}})$, where $y_{\texttt{q}} = \sum_{j} \omega_{\texttt{q},j} \cdot \delta_{\texttt{q},j}$, we aim to grade $y_{\texttt{q}}$ against empirically similar historical contracts in $\Dset$. The seller-facing output is a \textbf{letter grade} (A--F) derived from the \textbf{percentile rank} of $y_{\texttt{q}}$ within the peer discount distribution, together with \textbf{per-product-line sub-scores} that show which product-line discounts are generous relative to peers, giving directional guidance on where to adjust. The underlying \textbf{peer set} of similar contracts is available to the centralized review team and the model development team for audit and governance. The core difficulty is identifying peers in a feature space too rich for direct matching yet where the right similarity metric is not known in advance.

\subsection{Ensemble Trees as Adaptive Peer Finders}
\label{sec:method}

\begin{figure}[!h]
\centering
\includegraphics[width=0.75\columnwidth]{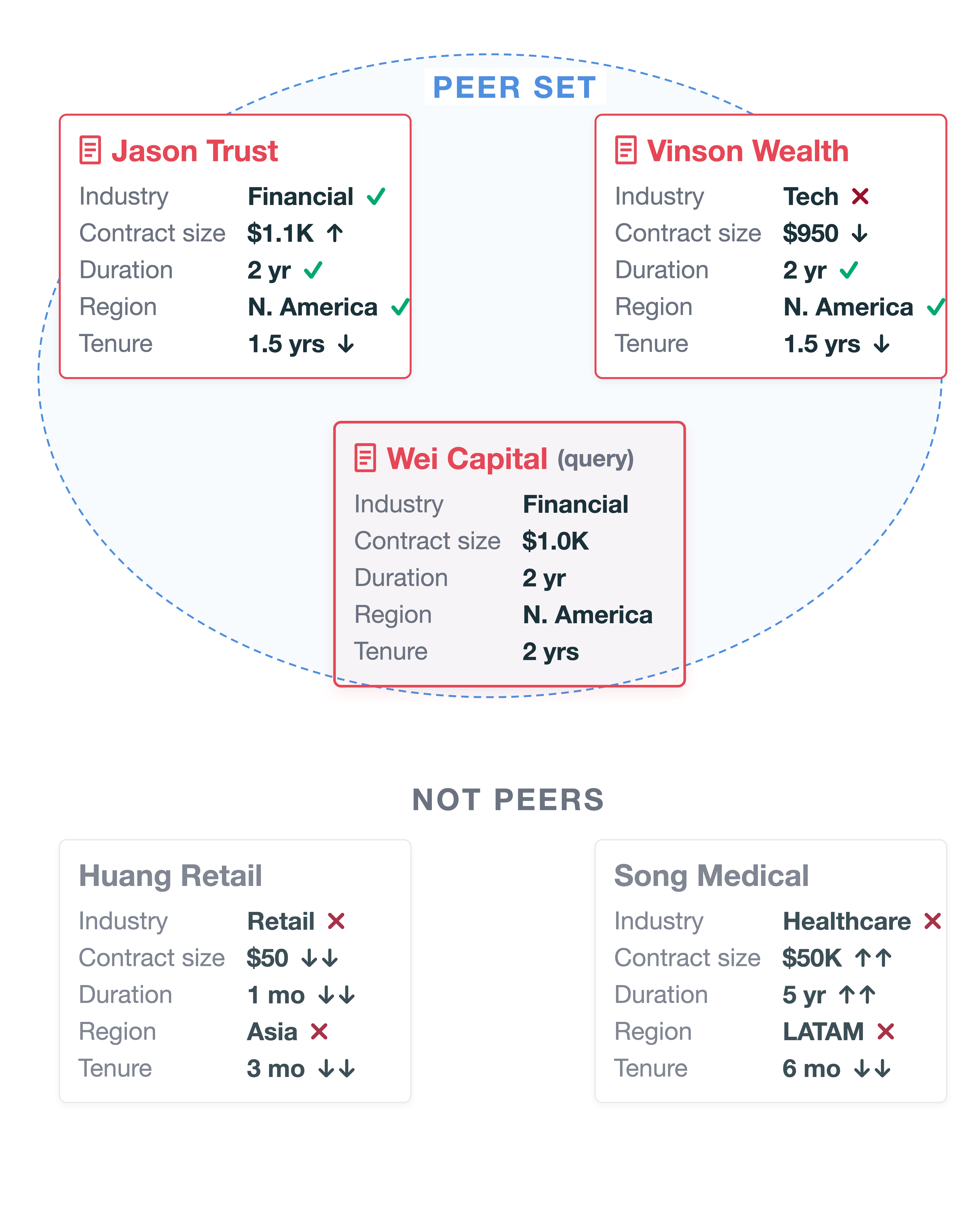}
\vspace{-0.35in}
\caption{Illustrative peer construction. Two peers (red) share leaf membership with the target across the ensemble; two non-peers (gray) differ sharply on the dimensions that explain discount variation.}
\label{fig:finding-peers}
\end{figure}

Ensemble trees address this directly: any ensemble tree estimator induces a data-driven similarity via shared leaf membership. \citet{LinJeon2006} showed that a random forest with $T$ trees produces predictions as a weighted average of training targets: $\hat{y}(x_{\texttt{q}}) = \sum_{i=1}^{N} w(x_{\texttt{q}}, x_i) \cdot y_i$, where the weights encode a similarity learned from the discount target. We do not use $\hat{y}$---we repurpose $w$ as a peer-retrieval primitive (Figure~\ref{fig:finding-peers}).

\vspace{-0.15in}
\vspace{-0.15in}

\begin{definition}[Leaf co-occurrence weight]\label{def:weight}
Let $\leaff_t(x)$ denote the leaf node of $x$ in tree $t$ and $L_t(x)$ the training contracts in that leaf. The weight is:
\begin{equation}
w(x_{\texttt{q}}, x_i) = \frac{1}{T} \sum_{t=1}^{T} \frac{\mathbf{1}[\leaff_t(x_i) = \leaff_t(x_{\texttt{q}})]}{|L_t(x_{\texttt{q}})|}.
\end{equation}
\end{definition}

The $1/|L_t|$ normalization gives higher weight to contracts in smaller leaves. The weights sum to one~\cite[Eq.~2.1]{LinJeon2006}, forming a proper distribution over training contracts.

\begin{definition}[Peer set]\label{def:peerset}
$P(x_{\texttt{q}}) = \{x_i \in \Dset : w(x_{\texttt{q}}, x_i) > 0\}$.
\end{definition}

The percentile rank of the proposed discount is:
\begin{equation}
r(y_{\texttt{q}}, P) = \frac{\sum_{x_i \in P:\, y_i < y_{\texttt{q}}} w(x_{\texttt{q}}, x_i)}{\sum_{x_i \in P} w(x_{\texttt{q}}, x_i)}.
\end{equation}
Since the weights sum to one~\cite[Eq.~2.1]{LinJeon2006}, the denominator equals one; we retain the fraction for parallel structure with per-product-line sub-score $r_j$'s, whose denominators could be less than one when peers lack a product line. Specifically, for product line $j$:
\begin{equation}
r_j(\delta_{\texttt{q},j}, P) = \frac{\sum_{x_i \in P\,:\,\delta_{i,j} < \delta_{\texttt{q},j}} w(x_{\texttt{q}}, x_i)}{\sum_{x_i \in P\,:\,\delta_{i,j} \neq \text{null}} w(x_{\texttt{q}}, x_i)},
\end{equation}
excluding peers without that product line; when no peers carry a given product line, $r_j$ falls back to a neutral value of $0.5$. Individual $r_j$ values are reported alongside $r$ for transparency.

\begin{definition}[Contract grade]\label{def:grade}
We define contract score as: $$g(x_{\texttt{q}}, y_{\texttt{q}}) = \alpha\,\bigl(1 - r(y_{\texttt{q}}, P)\bigr) + (1-\alpha)\,m_{\texttt{q}},$$ which translates into a letter grade (A--F) via fixed thresholds. The numeric score is internal; sellers see the letter grade and per-product-line breakdown. The full peer set is available to the centralized review team and model development team for audit. Here $m_{\texttt{q}}$ is a complementary signal capturing non-discount contract terms---such as free services including training and support---and $\alpha \in [0,1]$ controls the weight of the peer-based percentile relative to this signal; in production $\alpha > 0.5$, keeping the peer-based signal dominant.
\end{definition}

The complementary signal is deliberate: non-discount terms affect the overall value of the contract but are not captured by the discount percentile alone. Folding them into the grade ensures the score reflects the full contract, not just the discount.

Because the grade decomposes into per-product-line sub-scores and the endpoint responds in seconds, sellers can adjust specific discounts and re-query---using the scoring endpoint as a design API over the contract space. Three properties make this API tractable:

\begin{itemize}
    \item \emph{Monotonicity.} Lowering the discount on any product line weakly improves the grade. This is guaranteed by construction: the feature vector $x_{\texttt{q}}$ contains only non-discount contract attributes, so the peer set $P(x_{\texttt{q}})$ is invariant to discount changes. The usage-weighted discount $y_{\texttt{q}}$ is a convex combination of per-product-line discounts, so reducing any $\delta_{\texttt{q},j}$ reduces $y_{\texttt{q}}$, which decreases $r$ (fewer peers fall below the query) and therefore increases $(1-r)$. Because the peer set is fixed and the grade $g = \alpha(1-r) + (1-\alpha)\,m_{\texttt{q}}$ is monotonically increasing in $(1-r)$, the grade is non-decreasing in discount reductions---sellers can improve their grade only by lowering discounts, never by raising them.
    \item \emph{Decomposability.} Per-product-line sub-scores $r_j$ attribute grade movement to specific product lines, so sellers can identify which discount is dragging the grade down.
    \item \emph{Smoothness.} Small discount changes produce proportionate changes in the numeric score. The letter grade, however, is discrete; at grade thresholds sellers face a step change, creating the notches where bunching concentrates (\S\ref{sec:impact}).
\end{itemize}

\noindent For example, shifting 2~percentage points from a high-usage product line to a low-usage one can leave the customer's effective cost nearly unchanged while improving the contract's overall letter grade.

\section{Real World Impact}

A research-grade peer finder only matters if it runs on every contract at quote time---the moment a seller configures a proposed contract in the Configure-Price-Quote (CPQ) system, before the contract is signed---and measurably changes how sellers structure discounts. We describe the production pipeline and then present deployment evidence.

\subsection{Deployment}
\label{sec:pipeline}

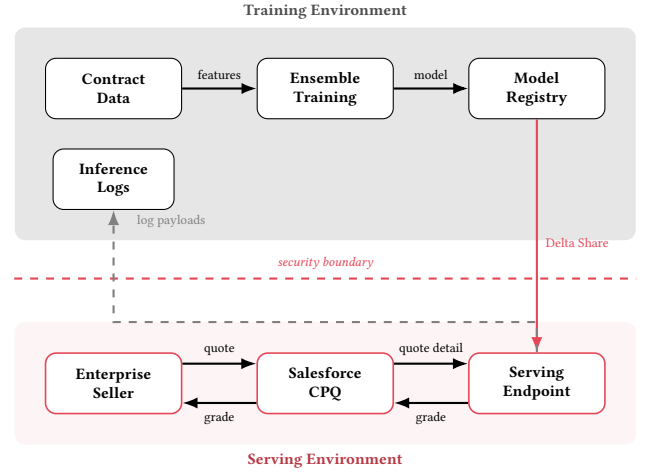
\begin{figure}[!h]
\centering
\begin{tikzpicture}[
  box/.style={draw, rounded corners, minimum height=8mm, minimum width=18mm,
              align=center, font=\scriptsize\bfseries, fill=white, line width=0.4pt},
  sbox/.style={box, draw=rulered, line width=0.6pt},
  flow/.style={-{Latex[length=2mm]}, thick},
  lbl/.style={font=\tiny},
  envfill/.style={rounded corners=4pt, inner sep=4mm}
]
\node[box] (data) at (0, 0) {Contract\\Data};
\node[box] (train) at (2.8, 0) {Ensemble\\Training};
\node[box] (reg) at (5.6, 0) {Model\\Registry};
\node[box, minimum width=16mm] (logs) at (0, -1.2) {Inference\\Logs};

\node[sbox] (ep) at (5.6, -3.9) {Serving\\Endpoint};
\node[sbox] (cpq) at (2.8, -3.9) {Salesforce\\CPQ};
\node[sbox] (seller) at (0, -3.9) {Enterprise\\Seller};

\begin{scope}[on background layer]
  \node[envfill, fill=black!10,
        fit=(data)(train)(reg)(logs),
        label={[font=\scriptsize\bfseries, text=black!70]above:Training Environment}] (tenv) {};
  \node[envfill, fill=rulered!6,
        fit=(ep)(cpq)(seller),
        label={[font=\scriptsize\bfseries, text=rulered!80!black]below:Serving Environment}] (senv) {};
\end{scope}

\draw[thick, dashed, rulered]
  ([yshift=-5mm]tenv.south west) --
  node[above, font=\tiny\itshape, text=rulered, pos=0.5]{security boundary}
  ([yshift=-5mm]tenv.south east);

\draw[flow] (data) -- node[above, lbl]{features} (train);
\draw[flow] (train) -- node[above, lbl]{model} (reg);

\draw[flow, rulered] (reg) -- node[right, lbl, text=rulered, pos=0.53]{Delta Share} (ep);

\draw[flow] ([yshift=2.6mm]cpq.east) -- node[above, lbl]{quote detail} ([yshift=2.6mm]ep.west);
\draw[flow] ([yshift=-2.6mm]ep.west) -- node[below, lbl]{grade} ([yshift=-2.6mm]cpq.east);
\draw[flow] ([yshift=2.6mm]seller.east) -- node[above, lbl]{quote} ([yshift=2.6mm]cpq.west);
\draw[flow] ([yshift=-2.6mm]cpq.west) -- node[below, lbl]{grade} ([yshift=-2.6mm]seller.east);

\draw[flow, dashed, black!50]
  (ep.north) -- ++(0, 0.4) -| (logs.south);
\node[lbl, black!50, anchor=west] at (0.2, -1.75) {log payloads};
\end{tikzpicture}
\vspace{-0.15in}
\caption{System architecture of Contract Scoring.}
\label{fig:architecture}
\end{figure}

As illustrated in Figure~\ref{fig:architecture}, historical contract records are joined to contract features, used to train an ensemble tree regressor; post-training, leaf assignments are extracted to reconstruct the peer weights at inference time. The model is shared via Delta Sharing~\cite{DeltaSharing2026} across a security boundary to a serving endpoint; Salesforce CPQ sends the quote detail and receives back a grade. \emph{The grade determines manual review intensity: lower grades trigger additional scrutiny.} Inference payloads are logged; periodic reviews of model performance and qualitative business review of peer quality inform periodic retraining decisions.

\vspace{-0.1in}
\vspace{-0.1in}

\subsection{Empirical Results}
\label{sec:impact}

\begin{figure}[!h]
\centering
\includegraphics[width=0.51\textwidth]{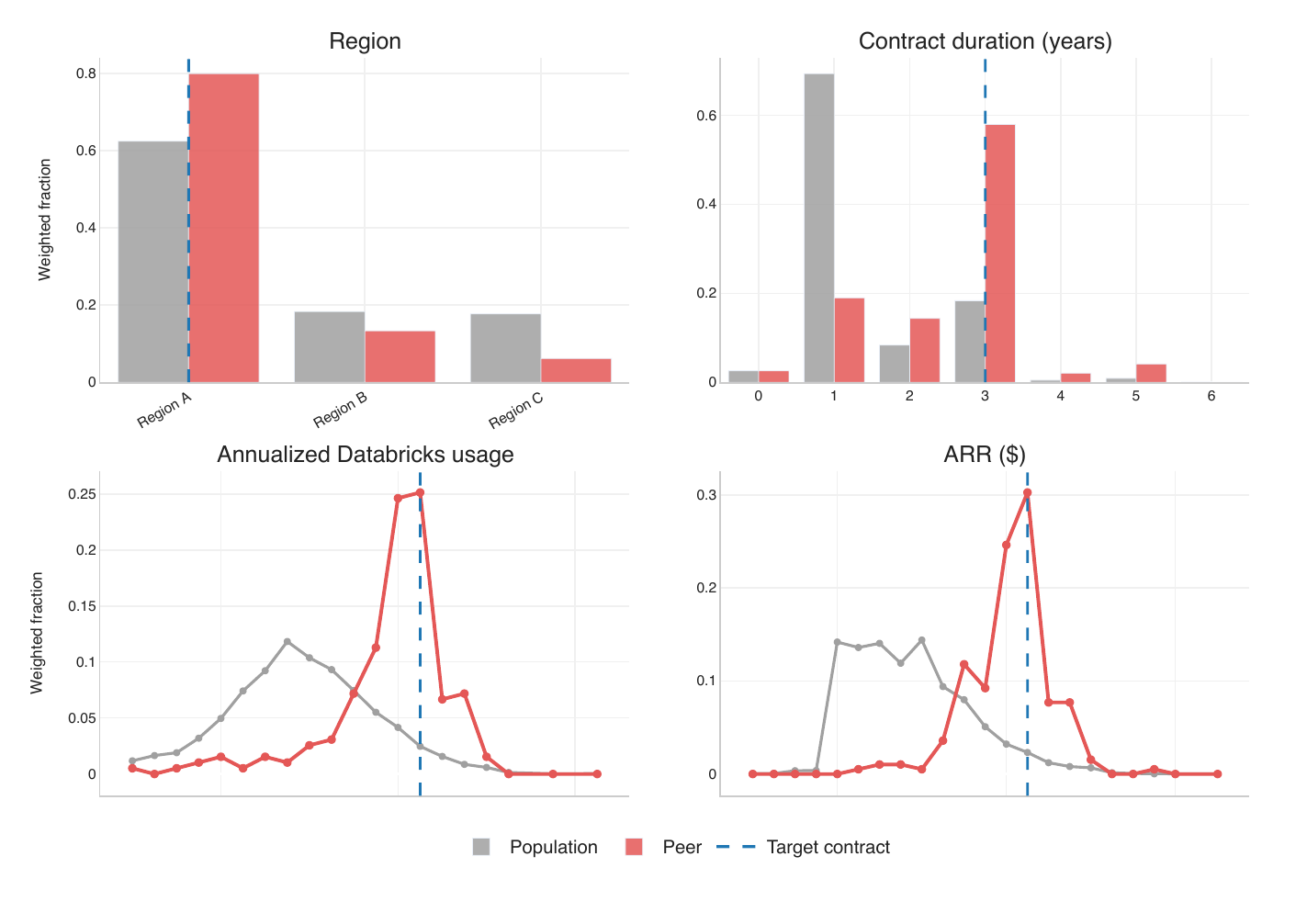}
\vspace{-0.4in}
\caption{Peer vs.\ population feature distributions for an illustrative hypothetical contract. The peer-weighted distribution (red) concentrates near the target on all four axes---even on Region, which the model is not given as input.}
\label{fig:peer-vs-pop}
\end{figure}

Contract Scoring has been deployed for the past year on every contract. The peer-quality evaluation shown in Figure~\ref{fig:peer-vs-pop} demonstrates the model is performing as expected: the peer set concentrates on the query's feature values. Region is the cleanest signal: the model is never given a region feature, yet the majority of peer mass lands in the target's region---consistent with the learned weights capturing discount-relevant similarity rather than a fixed proxy.

As a complementary check, the ensemble predicts the usage-weighted discount with cross-validated $R^2 \approx 0.6$, roughly double that of a model using only contract size and duration---evidence the richer feature set carries real discount signal. We report $R^2$ only as a sanity check on the learned similarity; the deployed system uses the peer weights $w$, not the point prediction~$\hat{y}$.

\vspace{-0.1in}
\vspace{-0.15in}

\begin{figure}[!h]
\centering
\includegraphics[width=0.9\columnwidth]{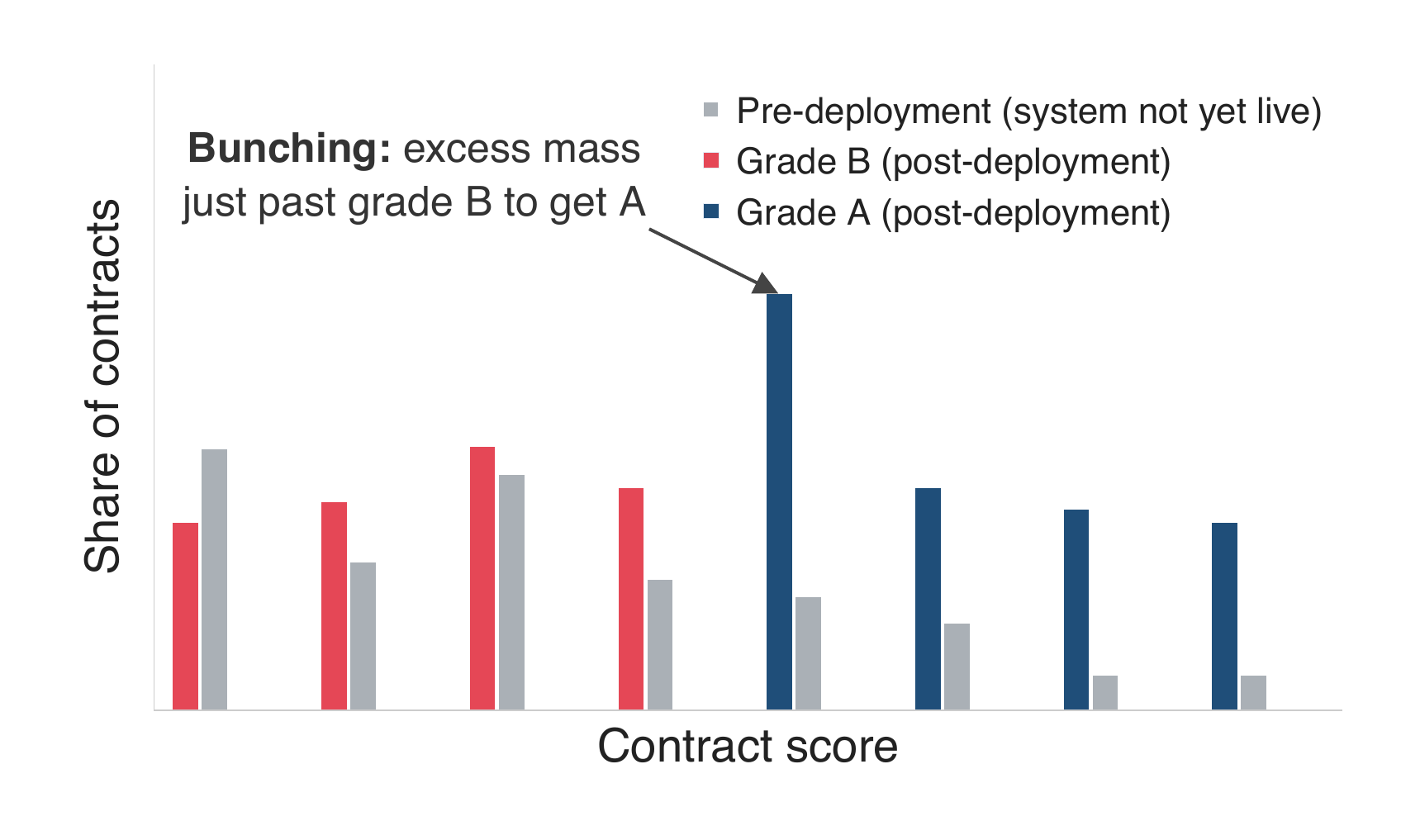}
\vspace{-0.25in}
\caption{Bunching at the B/A boundary. Colored bars: live contracts (red\,=\,B, blue\,=\,A). Gray bars: pre-deployment contracts (closed before the system existed) scored retroactively but never shown a grade. The pre-deployment baseline is smooth; the live distribution spikes just past the threshold.}
\label{fig:bunching}
\end{figure}

\noindent\paragraph{Real impact comes from the sellers ``behavioral change''.} Whether the grade moves discount-structuring behavior is an empirical question we must answer without a randomized control. We borrow the \emph{bunching} identification approach from public economics~\cite{Kleven2016}: when a continuous score is mapped to discrete grades at a threshold and agents care which side they land on, excess mass develops just above the cutoff.
Figure~\ref{fig:bunching} overlays two populations on the same grade axis. Colored bars are \emph{live} contracts graded post-deployment (the seller sees the grade); gray bars are \emph{pre-deployment} historical contracts---closed before the system existed---scored retroactively but never shown a grade, serving as a counterfactual for ``no behavioral response.'' The pre-deployment baseline declines smoothly; the live distribution spikes in the first A bucket, with a matching deficit immediately below the boundary.

\vspace{-0.15in}
\vspace{-0.1in}

Following \citet{Saez2010} and \citet{Kleven2016}, we partition the score domain into bins of width $h$ and fit a polynomial counterfactual $\hat{c}_\ell$ to bin counts outside a window around the threshold $b_{\text{thres}}$. The excess mass is:
\begin{equation}
\hat{B} = \frac{\sum_{\ell:\, b_\ell \in [b_{\text{thres}},\; b_{\text{thres}} + \Delta]} (c_\ell - \hat{c}_\ell)}{\bar{c}},
\end{equation}
where $c_\ell$ is the observed bin count and $\bar{c}$ the average counterfactual count. A positive $\hat{B}$ means more contracts land just above the threshold than a smooth distribution predicts. Standard errors are bootstrapped (Appendix~\ref{sec:app-estimator}). Multiplying the excess count by the mean annualized contract value yields a revenue estimate, pointing to \textbf{measurable discount discipline across the scored portfolio, with a commercially meaningful impact on revenue}.

\section{Conclusions}
\label{sec:conclusions}

Contract Scoring has been live for the past year on every contract at Databricks, returning an inspectable peer-based grade in seconds and driving measurable discount discipline across the scored portfolio, with a commercially meaningful impact on revenue. The contribution is not the model---adaptive $k$NN via ensemble trees is well-established---but the reframe: transparent, decomposable, real-time grading turns a governance gate into a design tool that sellers actively navigate. The bunching response at grade boundaries confirms that this design loop operates at scale.

\noindent\paragraph{Lessons learned.} Peer transparency, not model accuracy, drove adoption---the review team accepted the system because they could inspect and challenge the peer set, and sellers accepted the grade because the per-product-line breakdown made it actionable. A standard ensemble-tree primitive proved sufficient when the engineering effort went into the surrounding system (per-product-line decomposition, sub-second serving, Salesforce integration). And the design-API reframe was emergent: we built a governance tool and discovered it was a design API---the bunching evidence was the confirmation, not the intent.

\noindent\paragraph{Future directions.} Because sellers demonstrably respond to the grade, the scoring rubric---the weighting scheme that maps contract components to a single grade---becomes a policy lever. The production system grades more than the product-line discount: it bundles free services, complementary usage, one-time investments, and support into a single score, and the current weighting already drives measurable discount discipline. Different weightings encode different strategic priorities---emphasizing services if the margin leak is there, or product discount if that is the concern. A natural next step is to A/B test alternative rubrics on comparable seller groups to identify which weighting most effectively drives the desired behavioral shift. This reframes the system from a static governance tool to a tunable behavioral instrument.

\section*{Acknowledgments}
We are deeply grateful to Sam Shah, Feng Pan, and Divy Menghani from the Data Science leadership team at Databricks for their support, feedback, and guidance. We extend special thanks to Sam for his extensive help with detailed revisions and for facilitating the legal review process. We also greatly appreciate Ken Spencer and Shaun Tang from the Strategic Deal Pricing team, as well as our senior business leaders, Michael Kiermaier and Stephen Moss, for their close partnership, which made this impactful work possible.

\bibliographystyle{ACM-Reference-Format}
\bibliography{references}

\appendix


\section{Extended Literature Survey}
\label{sec:app-related-work}

No prior work combines peer-based contract grading with real-time design feedback in a deployed pricing system. Existing approaches each address part of the problem but fail on a specific axis: prescriptive pricing optimization~\cite{Colias2023,Biggs2021} requires demand-curve estimation and produces a black-box recommendation rather than a defensible peer comparison; comparable-firm valuation in finance~\cite{BhojrajLee2002,HobergPhillips2016} defines similarity by proxy rather than learning it from the discount target; and RF-based benchmarking~\cite{Kennedy2020} uses predictions, not leaf co-occurrence. Our system fills this gap by learning similarity directly from the target variable, using the peer set as the internal basis for grading and making it available to the review team for audit, and deploying at quote-time latency. A secondary technical contribution is the choice of RF over GBT for peer construction: GBT trees are sequential corrections whose leaf co-occurrence weights do not sum to one and lack a probabilistic interpretation, while RF weights sum to one by construction~\cite{LinJeon2006}---the normalization that makes our percentile computation well-defined.

\subsection{Adaptive $k$NN via Ensemble Trees}

\citet{LinJeon2006} established that random forest predictions can be written as a weighted average of training-set targets, where the weights arise from shared leaf membership---an adaptive nearest-neighbor scheme. The weight formula has a clean probabilistic interpretation: each training contract receives weight proportional to the fraction of trees in which it shares a leaf with the query, normalized so all weights sum to one. This normalization is the property we exploit for percentile computation---it makes the weights a proper distribution over training contracts. \citet{DaviesGhahramani2014} formalized the same structure as a random-forest kernel, connecting the leaf co-occurrence weights to the kernel methods.

\citet{Meinshausen2006} extended Lin and Jeon's weights to estimate the full conditional CDF via Quantile Regression Forests (QRF). Our weighted percentile rank is, in fact, Meinshausen's Equation~6 evaluated at the proposed discount---we are implicitly constructing the conditional CDF and reading off the percentile. The limitation, noted by \citet{Cevid2022}, is that the underlying splits are still optimized for mean prediction, not distributional accuracy. \citet{WagerAthey2018} and the subsequent Generalized Random Forests (GRF) framework of \citet{AtheyTibshiraniWager2019} further generalized the adaptive-weighting property to estimate arbitrary local moment conditions, providing asymptotic normality guarantees that Lin and Jeon lack. GRF could provide confidence intervals on our percentile ranks as a future extension. \citet{BiauScornet2016} provide a comprehensive survey situating these developments within the broader literature.

A natural alternative is gradient-boosted trees (GBT). \citet{He2014} established GBT leaf indices as a supervised feature transform for click-through rate prediction at Facebook, and GBT typically achieves lower prediction error than RF for mean estimation. However, GBT trees are sequential corrections---each fits residuals of previous trees---so leaf co-occurrence weights do not sum to one and lack the probabilistic interpretation that makes our percentile computation well-defined. This normalization gap is a concrete technical reason we use RF rather than GBT for peer-construction.

More broadly, these variants reveal that the tree's \emph{splitting criterion}---the objective by which it decides what makes two records similar---is itself a design axis. Standard CART splits reduce target variance; GRF~\citep{AtheyTibshiraniWager2019} tailors the split toward a causal target; Distributional Random Forests (DRF)~\cite{Cevid2022} adopt a target-free distributional criterion based on Maximum Mean Discrepancy. Because our system repurposes leaf co-occurrence weights for percentile computation rather than mean prediction, DRF's distributionally homogeneous leaves could produce more calibrated peer sets---we analyze this direction in detail in Appendix~\ref{sec:app-discussion}.

\subsection{Peer and Comparable Selection}

The problem of identifying empirically similar entities has a long history in finance, where ``comparable-firm'' valuation is a foundational methodology:
\begin{itemize}
    \item \citet{BhojrajLee2002} developed the warranted-multiple approach: project each firm into a one-dimensional score via OLS regression on fundamentals, then select peers as firms with the closest warranted multiple. The approach is structurally analogous to ours---fit a regression of the target on features, select peers by proximity in the output space---but the regression is linear, unable to capture the segment-specific interactions (e.g., high-ARR contracts in a specific industry having different discount dynamics) that our ensemble trees capture natively without requiring manual feature engineering or domain-specific distance metrics.
    \item \citet{LeeMaWang2015} took a behavioral approach, using SEC EDGAR co-search patterns as a similarity signal. Their co-search intensity is conceptually identical to our leaf co-occurrence weight: both measure how frequently two entities appear together in the same context---a search session for them, a tree leaf for us. The key difference is the source of co-occurrence: theirs comes from revealed investor behavior, which may reflect noisy sentiment or attention effects; ours comes from the tree structure trained on the discount target, so similarity is shaped by what matters for discount variation.
    \item \citet{GeertsemaLu2023} showed that gradient-boosted tree predictions can be decomposed into weighted averages of peer firm multiples, with the weights serving as a comparability measure. Working in equity valuation, they demonstrated that these ML-based multiples substantially outperform traditional models. Their weight decomposition is a related but distinct mechanism from our RF leaf co-occurrence weights (GBT weights do not sum to one), yet the shared principle---that tree-based models produce implicit peer weights---provides external validation for this class of approach. Kennedy et al.~\cite{Kennedy2020} are our closest methodological cousin, using random-forest-based benchmarking with peer groups for organizational performance evaluation. The key difference: they benchmark via RF predictions adjusted for contextual factors, whereas our work identifies the peer group itself via leaf co-occurrence and grades against the peer discount distribution directly.
    \item \citet{HobergPhillips2016} pioneered data-driven, firm-specific peer sets via NLP similarity on 10-K product descriptions. Their key insight---that each firm should have its own distinct set of competitors rather than being grouped by fixed industry codes---is the same principle behind our contract-specific peer sets. Their threshold-based peer selection (all firms above a minimum similarity) parallels our \autoref{def:peerset} (all contracts with non-zero leaf co-occurrence weight).
\end{itemize}

Across this literature, our contribution inverts the similarity source: where prior work defines similarity by a proxy (fundamentals, text, co-searches, predictions) with no guarantee that the returned peers match the query on the characteristics that matter for the target variable, we learn similarity directly from the target variable (usage-weighted discount) and verify feature-level peer match post hoc (Figure~\ref{fig:peer-vs-pop} in the main body of the paper).

\subsection{B2B Pricing and Revenue Management}

Revenue-management optimization~\cite{Bertsimas2017,MisicPerakis2020,Phillips2005} addresses pricing from a prescriptive perspective---find the optimal price or assortment given demand models and capacity constraints. The closest B2B pricing reference is \citet{Colias2023}, who use a hierarchical Bayes mixed logit to predict offer acceptance and optimize B2B product offers. Their integration with economic utility theory is principled: the mixed logit decomposes acceptance into product-feature sensitivities at the customer level, estimated via random coefficients that capture cross-segment heterogeneity. Two ideas could strengthen our system as future extensions. First, if we had sufficient contract-outcome data, a choice model could calibrate our grade thresholds to actual close rates---addressing the elasticity gap noted in our Limitations. Second, the mixed logit's random coefficients are an alternative to our tree-based segmentation that could provide distributional uncertainty estimates on discount sensitivity across segments.

\citet{Biggs2021} are motivated by the adoption barrier that interpretability poses for ML-driven pricing: complex models that perform well offline are rejected in practice because stakeholders cannot verify or trust how prices are formed. They propose model distillation into prescriptive trees as the solution. Our system serves the same function---the peer set makes the score defensible to the review team, who can inspect and challenge the comparison contracts---while sellers see only the grade and per-product-line breakdown, which provide the actionable signal without exposing contract-level detail.

\citet{Bernardi2019} validate at Booking.com that offline model accuracy does not necessarily translate to business value. Their central lesson---that the iterative hypothesis-driven process matters more than any single model improvement---directly motivates our deployment evidence section. We report bunching (behavioral response to the grade) rather than only $R^2$ precisely because the system's value lies in how sellers use the output, not in how accurately the model predicts the discount level.


\section{Bunching and Behavioral Economics}
\label{sec:app-estimator}

The bunching methodology we borrow originates in public economics. \citet{Saez2010} defined the foundational excess mass estimator for detecting behavioral responses at kink points in the tax schedule, showing that the amount of bunching at a kink is proportional to the compensated elasticity of income with respect to the net-of-tax rate. This proportionality result is why bunching is informative: the size of the spike tells you something about how strongly agents respond to the discrete incentive.

\citet{Kleven2016} provided a comprehensive survey and extended the methodology to notches (level changes in the choice set, as opposed to slope changes at kinks). The distinction is relevant for us: our grade thresholds create notches---a step change in the approval process and the seller's perceived scrutiny level---which the theory predicts produce sharper bunching than kinks. Kleven explicitly notes that bunching is finding applications beyond taxation, including private sector prices and reference-dependent preferences. Our application---detecting seller behavioral response to a discrete grade in enterprise contract pricing---falls squarely in this expanding frontier, and to our knowledge is the first application of the bunching estimator to B2B contract grading.

\paragraph{The Bunching Estimator.}
We follow the excess mass methodology of \citet{Saez2010} and \citet{Kleven2016}. Partition the score domain into equally-spaced bins $\{b_\ell\}_{\ell=1}^{J_b}$ of width $h$. Let $c_\ell$ denote the observed count of scored contracts in bin $b_\ell$, and let $b_{\text{thres}}$ denote the A/B grade boundary. The counterfactual density is estimated by fitting a polynomial to bin counts excluding a window $[b_{\text{thres}} - \Delta,\; b_{\text{thres}} + \Delta]$ around the threshold:
\begin{equation}
\hat{c}_\ell = \sum_{p=0}^{P} \gamma_p \cdot (b_\ell)^p, \quad \text{fitted on } \{\ell : b_\ell \notin [b_{\text{thres}} - \Delta,\; b_{\text{thres}} + \Delta]\},
\end{equation}
where the excess mass above the threshold is defined as:
\begin{equation}
\hat{B} = \frac{\sum_{\ell:\, b_\ell \in [b_{\text{thres}},\; b_{\text{thres}} + \Delta]} (c_\ell - \hat{c}_\ell)}{\bar{c}},
\end{equation}
where $\bar{c}$ is the average counterfactual bin count outside the exclusion window. A positive $\hat{B}$ indicates more contracts land just above the threshold than a smooth distribution would predict. The corresponding missing mass below the threshold---contracts ``pulled up'' by sellers adjusting their discount---is:
\begin{equation}
\hat{M} = \frac{\sum_{\ell:\, b_\ell \in [b_{\text{thres}} - \Delta,\; b_{\text{thres}})} (\hat{c}_\ell - c_\ell)}{\bar{c}}.
\end{equation}

Standard errors on $\hat{B}$ and $\hat{M}$ are bootstrapped by resampling contracts with replacement and re-estimating the counterfactual and excess mass per resample. The aggregate annual revenue impact is $\hat{R} = \hat{B} \cdot \bar{c} \cdot h \cdot \overline{\mathrm{ARR}}$: $\hat{B} \cdot \bar{c}$ is the total excess contract count in the bunching window (since $\hat{B}$ normalizes by $\bar{c}$, multiplying back recovers the raw count), $h$ (the bin width) bounds the per-contract discount reduction so that $h \cdot \overline{\mathrm{ARR}}$ estimates the revenue impact per bunching contract, where $\overline{\mathrm{ARR}}$ is the mean annual recurring revenue of contracts in the bunching region. The resulting estimate is commercially meaningful; we report the specific figure internally but not in this paper. The smooth pre-deployment distribution in the main-body figure is an independent, model-free check on the fitted counterfactual density~$\hat{c}_\ell$.

\noindent\paragraph{Interpreting the displacement pattern.} The bar immediately above the threshold is roughly twice as tall as the adjacent bars on either side, and the bucket immediately below shows a corresponding deficit---exactly the discontinuity the bunching framework predicts. If sellers treated the grade as a one-shot evaluation, we would expect no systematic clustering at the boundary---the discount distribution would reflect negotiation dynamics alone. Instead, the excess mass just above the boundary indicates that sellers adjust contract structures so the grade crosses it---consistent with using the grade as a design signal.

\noindent\paragraph{Ruling out an alternative mechanism.} An alternative explanation for the spike is that sellers target the maximum discount that still earns an A---high-A contracts drifting down to the margin---rather than B contracts being pulled up across the boundary. The displacement pattern rules this out: the matched missing mass sits just below the boundary (top of B), while the high-A region tracks the smooth pre-deployment baseline rather than showing the depletion a downward drift would produce. The excess at the margin is fed from below---sellers reducing discounts to cross into A.


\section{Extended Discussion}
\label{sec:app-discussion}

We build on a standard, well-understood ensemble-tree implementation rather than a specialized forest variant. Most of the engineering effort went into the surrounding system---model serving, per-product-line score decomposition, logging, and the Salesforce surface---and for a system graded on every contract, a predictable, stable primitive was worth more than a more sophisticated but less battle-tested splitting criterion. That said, the splitting criterion is the single most promising lever for methodological improvement.

\citet{Cevid2022} introduced Distributional Random Forests (DRF), which replace CART's variance-based split criterion with a Maximum Mean Discrepancy (MMD) two-sample test \citep{gretton2012kernel}. Where CART asks ``does this split reduce variance of $y$?'', DRF asks ``does this split maximize the distributional difference between child nodes?''---with a properly chosen kernel, DRF captures differences in variance, skewness, and modality, not just the mean.

\noindent\paragraph{Why this matters for us?} Our system trains a standard RF whose CART splits are optimized for mean prediction, but we never use the mean prediction $\hat{y}$. Instead, we repurpose the leaf co-occurrence weights for peer retrieval and compute the percentile rank, which depends on the full shape of the conditional distribution. This creates a mismatch:

\begin{tcolorbox}[colback=gray!10, colframe=gray!10, boxrule=0pt]
Consider two candidate splits at a tree node:
\begin{itemize}
    \item Split~A separates contracts into children with mean discounts 15\% vs.\ 25\% (large mean difference, high CART score) but both children have broad, overlapping distributions.
    \item Split~B produces children with means 19\% vs.\ 21\% (small mean difference, low CART score) but one child is tightly concentrated while the other is bimodal.
\end{itemize}
CART chooses Split~A; for percentile computation, Split~B is superior because the distributionally distinct leaves produce more calibrated peer sets.
\end{tcolorbox}

\noindent\paragraph{MMD formulation.} Given a kernel $\kappa$ (e.g., DRF uses a Gaussian kernel), the MMD between child-node distributions is:
\begin{align*}
\mathrm{MMD}^2(\mathcal{C}_L, \mathcal{C}_R)
  &= \frac{1}{|\mathcal{C}_L|^2}\sum_{i,j \in \mathcal{C}_L} \kappa(y_i, y_j)
   + \frac{1}{|\mathcal{C}_R|^2}\sum_{i,j \in \mathcal{C}_R} \kappa(y_i, y_j) \\
  &\quad - \frac{2}{|\mathcal{C}_L||\mathcal{C}_R|}\sum_{\substack{i \in \mathcal{C}_L \\ j \in \mathcal{C}_R}} \kappa(y_i, y_j).
\end{align*}
DRF replaces CART's criterion with this statistic at every split, producing leaves that are more distributionally homogeneous.

As part of our ongoing development, the real-world deployment will feature the following novel adaptations:
\begin{itemize}
    \item DRF natively handles multivariate responses. Instead of training on the scalar usage-weighted discount $y_i$, we could train on the vector $(\delta_{i,1}, \ldots, \delta_{i,J})$ of per-product-line discounts directly. The MMD criterion would then capture correlation structure across product lines---for example, contracts where two product-line discounts co-move vs.\ contracts where they are independent---allowing per-product-line sub-scores to emerge from a single jointly-optimized model rather than being derived post hoc.
    \item The MMD computation is $O(n_{\text{node}}^2)$ per split candidate vs.\ $O(n_{\text{node}})$ for CART, though DRF mitigates this via subsampling; for our training set of thousands of contracts this is likely acceptable, or we could adopt the linear-time MMD \citep{gretton2012kernel} or block MMD \citep{zaremba2013b,li2015m,wei2026online} to trade statistical power for computing efficiency.
    \item The Gaussian kernel bandwidth $\sigma$ could be set via median heuristic and would need validation on our discount distributions. Critically, the right evaluation metric for the upgrade is not $R^2$ but \emph{percentile calibration}: for contracts whose peer-set percentile is $p$, do approximately $p\%$ of held-out peers actually have lower discounts? Our current RF achieves cross-validated $R^2 \approx 0.6$ vs.\ the size-plus-duration baseline with $R^2 \approx 0.3$---the DRF upgrade could be principled but not yet operationally urgent.
\end{itemize}

\end{document}